\documentclass[conference]{IEEEtran}  

\usepackage[utf8]{inputenc}
\usepackage[english]{babel}
\usepackage[T1]{fontenc} 

\usepackage[cmex10]{amsmath}
\usepackage{amsfonts,amsthm,amssymb}

\usepackage{xspace} 
\usepackage{datetime} 
\usepackage{soul} 
\usepackage{cite} 

\usepackage{marginnote}

\usepackage{tikz}
\usetikzlibrary{decorations.pathreplacing}

\renewcommand{\ge}{\geqslant}
\renewcommand{\le}{\leqslant}

\newcommand{\ff}{\mathbb{F}}
\newcommand{\cS}{\mathcal{S}}

\newcommand{\functor}[1]{\ensuremath{\mathsf{#1}}}

\newcommand{\prob}{\functor{P}\xspace}
\newcommand{\EE}{\functor{E}\xspace}

\newcommand{\cw}[1]{\ensuremath{\mathbf{#1}}\xspace}
\newcommand{\tr}{^\intercal\xspace}
\newcommand{\rank}{\operatorname{rank}}

\newcommand{\code}{\ensuremath{\mathcal C}\xspace}
\newcommand{\dual}{\ensuremath{\mathcal{C}^\perp}\xspace}
\newcommand{\dualz}{\ensuremath{\mathcal{C}^\perp_0}\xspace}

\newcommand{\Nat}{\ensuremath{\mathbb N}\xspace}

\newtheorem{theorem}{Theorem}
\newtheorem{lemma}{Lemma}
\newtheorem{cor}{Corollary}
\newtheorem{define}{Definition}

\title{Refined Upper Bounds on Stopping Redundancy of Binary Linear Codes
} 

\author{{\bf Yauhen Yakimenka, Vitaly Skachek}\\
	Institute of Computer Science\\
	University of Tartu, Estonia\\
	Emails: {\tt \{ yauhen, vitaly \} @ut.ee}} 
	
\begin{document}
\maketitle

\begin{abstract}
The $l$-th stopping redundancy $\rho_l(\code)$ of the binary $[n, k, d]$ code $\code$, 
$1 \le l \le d$, is defined as the minimum number of rows in the parity-check matrix 
of $\code$, such that the smallest stopping set is of size at least $l$. 
The stopping redundancy $\rho(\code)$ is defined as $\rho_d(\code)$.
In this work, we improve on the probabilistic analysis of stopping redundancy, 
proposed by Han, Siegel and Vardy, which yields the best bounds known today. 
In our approach, we judiciously select the first few rows in the parity-check matrix, 
and then continue with the probabilistic method. By using similar techniques, we 
improve also on the best known bounds on $\rho_l(\code)$, for $1 \le l \le d$. 
Our approach is compared to the existing methods by numerical computations.  
\end{abstract}



\renewcommand{\thefootnote}{\fnsymbol{footnote}}
\footnotetext{This work is supported by the Norwegian-Estonian Research Cooperation Programme under the grant EMP133, by the Estonian Ministry of Education and Research through the research grants PUT405 and IUT2-1, and by the European Regional Development Fund through
the Estonian Center of Excellence in Computer Science, EXCS. The work of the first author is also supported by the HITSA Tiger University programme.} 
\renewcommand{\thefootnote}{\arabic{footnote}}

\section{Introduction}

\emph{Stopping sets} are a known cause of failures of message-passing decoders, when applied to binary linear codes on a binary erasure 
channel~\cite{di2002finite}. Small stopping sets are especially harmful, as they have higher probability of causing the damage. 
Stopping sets, however, are determined by the selection of a parity-check matrix of the code, rather than by the code itself. 
The size of the smallest stopping set is called the \emph{stopping distance} of the corresponding parity-check matrix.  

It is observed in~\cite{santhi2004effect} that by adding redundant rows to the parity-check matrix, the small stopping sets can be eliminated, 
i.e. the resulting matrix does not contain stopping sets of small size.  
On the other hand, the increased number of the redundant rows in the parity-check matrix leads to growth in the decoding complexity. Therefore, generally, the trade-off between the size of the smallest stopping set, and the number of rows in the parity-check matrix, is of significant interest. 

More specifically, let $\code$ be a binary linear $[n, k, d]$ code, and let $H$ be a parity-check matrix for this code. 
Denote $[n] \triangleq \{ 1,2,\dotsc,n \}$. Let $\cS \subseteq [n]$ be a set of columns of $H$. Denote by $H_{\cS}$ the submatrix of $H$, composed from the columns of $H$ indexed by $\cS$. 

\begin{define} 
The set $\cS$ is a \emph{stopping set} in $H$ if $H_{\cS}$ contains no row of Hamming weight one. 
\end{define}  

\begin{define}[\hspace{-0.6ex} \cite{schwartz-vardy2006}] 
The stopping redundancy of $\code$, $\rho(\code)$, is the smallest number of rows in any parity-check matrix of $\code$, such that the corresponding stopping distance is $d$. 
\label{def:stopping-redundancy}
\end{define}  

Bounds on stopping redundancy of binary linear codes were studied in a number of works over the years~\cite{schwartz-vardy2006, weber2005stopping, etzion2006stopping, han2007improved, hollmann2007parity, olgica2008permutation, han2008improved, zumbraegel2012pseudocodeword}. Algorithms for finding small stopping sets were proposed in~\cite{rosnes2009efficient, karimi2013efficient}. 

For general binary linear codes, the best known bounds on the stopping redundancy were derived by using probabilistic 
method in~\cite{han2008improved}. In this work, we improve on the analysis therein. In particular, we observe that the number of stopping sets eliminated by a random codeword of the dual code is not optimal in general case. 
In our approach, we judiciously select the first few rows in the parity-check matrix, in such way that these rows eliminate more small stopping sets than the randomly chosen nonzero codewords in the dual code. In particular, we pick dual codewords of the minimum weight.   
If the number of such codewords  is small (for example, 1 or 2), then we can provide good estimates on the number of eliminated stopping sets. After that, we proceed with the probabilistic method, similarly to~\cite{han2008improved}.


\section{General Theorem}
\label{sec:gen-thm}
Throughout the remaining sections, if not explicitly stated otherwise, we consider a binary linear $[n, k, d]$ code \code. As it was shown in~\cite[Theorem~3]{schwartz-vardy2006}, if $d \le 3$ then \textit{any} parity-check matrix $H$ for \code has stopping distance $d$, 
i.e. $\rho(\code) = n-k$. Hence we only consider a case $d \ge 4$ (and, therefore, $r \triangleq n-k \ge 2$).

The dual code of \code is denoted by \dual, its dimension and minimum distance are $r$ and $d^\perp$, respectively. We use \dualz as a shorthand for $\dual \setminus \{ \cw 0 \}$.

We call any subset of $[n]$ of cardinality $i$ an \emph{$i$-set}. The set of all $i$-sets is denoted by $\mathfrak I_i$:
\[
	\mathfrak I_i = \{ \mathcal S \subseteq [n] : |\mathcal S| = i\} \; . 
\]

We also use the notation
$\mathfrak{I} = \bigcup_{i=3}^{d-1} \mathfrak{I}_i$.
We do not consider the $i$-sets of sizes $1$ and $2$. Indeed, if $d \ge 4$ then no parity-check matrix has the all-zero column or two identical columns, which implies there are no stopping sets of sizes $1$ and $2$.

We say that a row vector $\cw h \in \mathbb F_2^n$ \emph{covers} the $i$-set $\mathcal{S}$ if the projection of $\cw h$ on the coordinates indexed by $\mathcal{S}$ has Hamming weight $1$. We also say that the $t \times n$ matrix $(\cw h_1\tr,\cw h_2\tr,\dotsc,\cw h_t\tr)\tr$ over $\ff_2$ \emph{covers} $\mathcal{S}$ if any of its rows covers $\mathcal{S}$. If some $i$-set is covered, then the stopping set in the corresponding 
coordinates cannot exist. Thus, by covering all the $i$-sets, $i=3,4,\dotsc,d-1$, we obtain a matrix with no stopping sets of size less than~$d$.

The following lemma is implicitly stated in~\cite{han2008improved}. 


\begin{lemma}
\label{lm:b-lemma}
Let $r \ge 3$ and $d$ be two positive integers, and $b$ be a real number, such that $1 \le b \le r-2$, and 
$(r-1)(d-1) \le 2^{d-1}$. 
Then, for any $x<2^r$, 
\[
    b - \left( \frac{2^r-2^{r-b}}{2^r-x} \right)
    \le
    b \left( 1 - \frac{(d-1) \cdot 2^{r-d+1}}{2^r-x} \right) \; . 
\]
\end{lemma}

We omit the proof of Lemma~\ref{lm:b-lemma}. Next, we formulate a general theorem, which is the main result of this paper. It includes Theorem 7 in~\cite{han2008improved} as a special case, and its proof uses similar ideas. 

\begin{theorem}
\label{thm:general-upper-bound}
Assume that there exists a matrix, whose rows  $\cw h_1, \cw h_2, \dotsc, \cw h_\tau$, $\tau \ge 0$, are 
linearly independent codewords in \dualz. 
For $i=3,4,\dotsc,d-1$, let $\mathfrak U_i$, $|\mathfrak U_i| \le u_i$, be the set of $i$-sets not covered by this matrix. Assume also that $(r-1)(d-1) \le 2^{d-1}$. Then
\begin{equation}
\rho(\code) \le \tau + \min_{t \ge r} \left\{ t + \kappa_t  \right\},
\end{equation}
where
\begin{eqnarray*}
\kappa_t & = & \min \left\{k \in \Nat : Q_k(\lfloor \mathcal D_t \rfloor) = 0\right\} \; , \\
Q_k(x) & = & P_k(P_{k-1}(\ldots P_1(x) \ldots)) \; , \\
P_j(x) & = & \left\lfloor x \left( 1 - \frac{(d-1) \cdot 2^{r-d+1}}{2^r-(\tau+t+j)} \right) \right\rfloor \; , \\
\mathcal D_t & = & \sum_{i=3}^{d-1} u_i \prod_{j=\tau+1}^{\tau+t}\left(1 - \frac{i \cdot 2^{r-i}}{2^r-j}\right) \\ 
&& \qquad + \; \frac{1}{2^{t-r}}\left(1+\frac{2/3}{2^{t-r+1}-1}\right) \; . 
\end{eqnarray*}
\end{theorem}
\begin{IEEEproof}
Let $H$ be a matrix with rows in $\dualz$. Such $H$ is not necessary the parity-check matrix, since its rank can be less than $r$. Define $\delta(H)$ as follows:
\begin{align*}
    \delta(H) \triangleq \Big| \{\mathcal{S} \in \mathfrak{I} ~|~ \mathcal{S} &\mbox{ is not covered by } H \} \Big| 
        + (r - \rank H) . 
\end{align*}

Here $\delta(H) = 0$ means that $\rank H = r$ and all the $i$-sets, $i=3,4,\dotsc,d-1$, are covered. Such $H$ is a parity-check matrix of $\code$, and since its stopping distance is at least $4$, all the $1$-sets and $2$-sets are covered automatically. In the sequel, we construct a matrix $H$, such that $\delta(H) = 0$.

We prove this theorem in two steps. First, we show existence of a parity-check matrix of size $(\tau+t) \times n$ with bounded $\delta$. Second, we show that $\delta$ has to decrease after adding one carefully selected additional row to it. Therefore, after adding enough rows, we obtain a parity-check matrix $H$ with $\delta(H) = 0$. Hereafter, we use $H_{i_1,i_2,\dotsc,i_s}$ as a shorthand for the matrix with rows $\cw h_{i_1}, \cw h_{i_2}, \dotsc, \cw h_{i_s}$.

\textit{Step 1}. Let $\cw h_{\tau+1}, \cw h_{\tau+2}, \dotsc, \cw h_{\tau+t}$ be $t$ rows drawn uniformly at random without repetitions from $\dualz \setminus \{\cw{h}_1, \cw{h}_2, \dotsc, \cw{h}_\tau\}$. Denote by $\xi$ the number of sets in $\mathfrak I$ that are not covered by $H_{1,2,\dotsc,\tau+t}$. This $\xi$ is an integer discrete random variable. Denote by $\functor I \{\cdot\}$ an indicator function, which takes values $0$ and $1$. The value of the indicator is set to $1$ if the argument is true, and zero otherwise. Then, $\xi$ can be written as follows.
\begin{align*}
    \xi &= \sum_{\mathcal S \in \mathfrak I} \functor{I}\{\mathcal{S} \mbox{ is not covered by } H_{1,2,\dotsc,\tau+t}\} \\
    &= \sum_{i=3}^{d-1}\sum_{\mathcal{S} \in \mathfrak U_i} \functor{I}\{\mathcal{S} \mbox{ is not covered by } H_{\tau+1,\tau+2,\dotsc,\tau+t}\} \; .
\end{align*}
Then, the expected value of $\xi$ is
\begin{multline}
\label{eq:E_xi_for_pasting}
    \sum_{i=3}^{d-1} \sum_{\mathcal S \in \mathfrak U_i} \prob 
		\left\{ \mathcal S \mbox{ is not covered by }H_{\tau+1,\tau+2,\dotsc,\tau+t} \right\} \; .
\end{multline}

To find the probabilities in~(\ref{eq:E_xi_for_pasting}), recall (cf.~\cite[p.~139]{macwilliams1977theory}) that $2^r \times n$ matrix, consisting of all codewords of \dual, is an orthogonal array of strength $d-1$. This means that for any $i=3,4, \dotsc, d-1$, the projection of this matrix on any $i$-set $\mathcal{S}$ contains every vector of length $i$ exactly $2^{r-i}$ times. There are exactly $i \cdot 2^{r-i}$ codewords in $\dual_0$ that cover $\mathcal{S}$. Therefore, 
\begin{multline}
    \prob \left\{ \mathcal{S} \mbox{ is not covered by } H_{\tau+1,\tau+2,\dotsc,\tau+t} \right\}
    \\
    = \left.
    \binom{(2^r - \tau - 1) - i \cdot 2^{r-i}}{t}
    \right/
    {\binom{2^r - \tau - 1}{t}} \\
    = \prod_{j=\tau+1}^{\tau+t} \left( 1 - \frac{i \cdot 2^{r-i}}{2^r-j} \right) \; .
		\label{eq:prob-set-covered}
\end{multline}

In a numerator we have a number of possible choices of $\cw h_{\tau+1}, \cw h_{\tau+2}, \dotsc, \cw h_{\tau+t}$ that do not cover $\mathcal S$,  and in a denominator -- the total number of choices of $\cw h_{\tau+1}, \cw h_{\tau+2}, \dotsc, \cw h_{\tau+t}$.

By substituting expression~(\ref{eq:prob-set-covered}) into~(\ref{eq:E_xi_for_pasting}) we have that the expected value of $\xi$ is bounded from above by:
\begin{equation}
\label{eq:E_xi_bound}
\EE\{\xi\} \le \sum_{i=3}^{d-1} u_i \prod_{j=\tau+1}^{\tau+t} \left( 1-\frac{i \cdot 2^{r-i}}{2^r-j} \right) \; . 
\end{equation}

Next, it was shown in \cite[Lemma 6]{han2008improved} that if we draw uniformly at random $s$ codewords from \dual, $s \ge r$, then the matrix constructed from these codewords has expected rank at least
\begin{equation*}
r - \frac{1}{2^{s-r}}\left( 1+\frac{2/3}{2^{s-r+1}-1} \right) \; . 
\end{equation*}
It is easy to see that if we draw $\cw h_{\tau+1}, \cw h_{\tau+2}, \dotsc, \cw h_{\tau+t}$ uniformly at random from $\dualz \setminus \{\cw{h}_1, \cw{h}_2, \dotsc, \cw{h}_\tau\}$, and then construct the matrix $H_{1,2,\dotsc,\tau+t}$, 
then the expected value of its rank deficiency is bounded from above: 
\begin{eqnarray}
\EE \{ \eta \} & = & r - \EE \{ \rank H_{1,2,\dotsc,\tau+t} \} \nonumber \\
& \le & \frac{1}{2^{t-r}}\left( 1+\frac{2/3}{2^{t-r+1}-1} \right) \; .
\label{eq:E_eta_bound}
\end{eqnarray}

By summing up (\ref{eq:E_xi_bound}) and (\ref{eq:E_eta_bound}), we obtain that
\begin{multline*}
\EE \{ \delta(H_{1,2,\dotsc,\tau+t}) \} 
\; \le \; \sum_{i=3}^{d-1} u_i \prod_{j=\tau+1}^{\tau+t} \left( 1-\frac{i \cdot 2^{r-i}}{2^r-j} \right) \\
+ \frac{1}{2^{t-r}}\left( 1+\frac{2/3}{2^{t-r+1}-1} \right) \; . 
\end{multline*}

Since $\delta(H_{1,2,\dotsc,\tau+t})$ is an integer discrete random variable, there is a realisation of it such that
\begin{multline*}
\delta(H_{1,2,\dotsc,\tau+t}) \le \Bigg\lfloor \sum_{i=3}^{d-1} u_i \prod_{j=\tau+1}^{\tau+t} \left( 1-\frac{i \cdot 2^{r-i}}{2^r-j} \right)  \\
+ \frac{1}{2^{t-r}}\left( 1+\frac{2/3}{2^{t-r+1}-1} \right) \Bigg\rfloor \; . 
\end{multline*}

\textit{Step 2}. At this point we consider $\cw h_1, \cw h_2, \dotsc, \cw h_{\tau+t}$ as non-random and fixed. In particular, $\xi$ and $\eta$ are non-random. Let $\mathfrak U \subset \mathfrak I$ be the set of all $i$-sets ($3 \le i \le d-1$) not covered by $H_{1,2,\dotsc,\tau+t}$. Add one more new row $\cw h_{\tau+t+1}$, which is randomly chosen from $\dualz \setminus \{ \cw h_1, \cw h_2, \dotsc, \cw h_{\tau+t} \}$. Analogously to $\xi$ and $\eta$ for $H_{1,2,\dotsc,\tau+t}$, we define discrete random variables $\xi'$ and $\eta'$ for $H_{1,2,\dotsc,\tau+t+1}$. Then,
\begin{eqnarray*}
\EE \{ \xi' \} & = & \sum_{\mathcal S \in \mathfrak U} \prob \{ \mathcal S \mbox{ is not covered by } H_{1,2,\dotsc,\tau+t+1}\} \\
& \le &  |\mathfrak U| \cdot \max_{\mathcal S \in \mathfrak U} \prob \{ \mathcal S \mbox{ is not covered by } \cw h_{\tau+t+1} \} \\
& = & \xi \cdot \max_{\mathcal S \in \mathfrak U} \left( 1 - \frac{|\mathcal S| \cdot 2^{r-|\mathcal S|}}{2^r - (\tau+t+1)} \right) \\
& \le & \xi \left( 1 - \frac{(d-1) \cdot 2^{r-d+1}}{2^r - (\tau+t+1)} \right) \; . 
\end{eqnarray*}

Adding one row to any matrix could either leave its rank unchanged or increase it by one. Therefore, if $\eta \ge 1$ then\footnote{Note that the case $\eta \ge 1$ is possible only for $r \ge 3$.} we have that either $\eta' = \eta$ or $\eta' = \eta-1$. To calculate the probabilities of these events, we note that any $l$ linearly independent rows in \dualz span in total $2^l$ codewords (including $\cw 0$). Then 
\begin{equation*}
\prob \{ \eta' = \eta \} = \frac{2^{r-\eta} - (\tau+t+1)}{2^r - (\tau+t+1)} = 1 - \prob \{ \eta' = \eta-1 \} \; ,
\end{equation*}
and, therefore,
\begin{eqnarray*}
\EE \{ \eta' \} &=& \eta - \left( \frac{2^r - 2^{r-\eta}}{2^r - (\tau+t+1)} \right) \; .
\end{eqnarray*}

Next, apply Lemma \ref{lm:b-lemma} with $b = \eta$ and $x = \tau+t+1$. Indeed, $\eta \ge 1$ and $\eta \le r-2$ because $H_{1,2,\dotsc,\tau+t}$ consists of at least two different non-zero codewords. Additionally, $\tau+t+1 < 2^r$ since $2^r-1$ is the maximum number of rows in any parity-check matrix for \code. Therefore,
\begin{equation}
\EE \{ \eta' \}  \le \eta \left( 1 - \frac{(d-1) \cdot 2^{r-d+1}}{2^r - (\tau+t+1)} \right) \; . 
\label{eq:eta-prime}
\end{equation}

Inequality~(\ref{eq:eta-prime}) holds also when $\eta = 0$ (which includes the case $r=2$), because in that case $\eta' = 0$ as well.

Altogether we have
\begin{multline*}
\EE \{ \delta(H_{1,2,\dotsc,\tau+t+1}) \} = \EE\{ \xi' \} + \EE\{ \eta' \} \\
\le \delta(H_{1,2,\dotsc,\tau+t}) \left( 1 - \frac{(d-1) \cdot 2^{r-d+1}}{2^r - (\tau+t+1)} \right) \; . 
\end{multline*}

Therefore, there exists $\cw h_{\tau+t+1}$ such that $\delta(H_{1,2,\dotsc,\tau+t+1}) \le P_1(\delta(H_{1,2,\dotsc,\tau+t})) \le P_1(\lfloor \mathcal D_t \rfloor)$. We iterate this process of adding rows one-by-one, and after $k$ steps obtain the $(\tau+t+k) \times n$ matrix $H_{1,2,\dotsc,\tau+t+1}$ with $\delta(H_{1,2,\dotsc,\tau+t+1}) \le Q_k(\lfloor \mathcal D_t \rfloor)$.

Iterations should be stopped when $Q_k(\lfloor \mathcal D_t \rfloor) = 0$.
\end{IEEEproof}


\section{Important Special Cases}
\label{sec:special-cases}

Theorem~\ref{thm:general-upper-bound} gives a general family of bounds on the stopping redundancy. 
It remains a question how to choose particular $\tau$ and $\cw h_1, \cw h_2, \dotsc, \cw h_\tau$, 
which yield good concrete bounds. In this section, we study specific selections of these  
parameters. 

The first and simple choice is to take $\tau = 1$ and $\cw h_1$ to be a fixed codeword of the minimum weight 
in $d^\perp$.

\begin{cor}
\label{cor:1-row}
The upper bound in Theorem \ref{thm:general-upper-bound} holds for $\tau=1$ and 
\[
	u_i = \binom{n}{i} - d^\perp \binom{n-d^\perp}{i-1} \; \mbox{ for } i = 3, 4, \dotsc, d-1 \; . 
\]
\end{cor}
\begin{IEEEproof}
Matrix consisting of one codeword of weight $d^\perp$ covers exactly $d^\perp \binom{n-d^\perp}{i-1}$
$i$-sets for each $i = 3,4,\dotsc,d-1$. We apply Theorem \ref{thm:general-upper-bound} with $\tau=1$ and $u_i = \binom{n}{i} - d^\perp \binom{n-d^\perp}{i-1}$, which yields the result stated in the corollary.
\end{IEEEproof}

Next, take $\tau = 2$ and consider two different codewords of weight $d^\perp$.

\begin{cor}
\label{cor:2-rows}
If there are at least two different codewords $\cw h_1, \cw h_2 \in \dual$ of weight $d^\perp$, then the upper bound in Theorem~\ref{thm:general-upper-bound} 
holds for $\tau=2$, where
\begin{eqnarray*}
&& u_i \; = \; \binom{n}{i} - \mathfrak{M}(n, d^\perp, i) \; , \\
&&\mathfrak{M}(n, d^\perp, i) \triangleq 2 d^\perp \binom{n-d^\perp}{i-1} \; - \; \max_{0 \le \Delta \le \lfloor d^\perp / 2 \rfloor} \Bigg\{ \Delta \cdot \\
&& \quad \binom{n-2 d^\perp+\Delta}{i-1} \; + \; (\Delta - d^\perp)^2 \binom{n-2 d^\perp+\Delta}{i-2} \Bigg\} \; . \\
\end{eqnarray*}

\end{cor}
\begin{IEEEproof}
Consider two different codewords in \dualz of weight $d^\perp$. They are shown in Figure \ref{fig:2-codewords}, where grey and white colors denote the regions of ones and zeroes, respectively. Let $\Delta$ be the number of codeword positions, where both of the codewords have ones. Obviously $0 \le \Delta \le \lfloor d^\perp / 2 \rfloor$.

\begin{figure}
\centering
\begin{tikzpicture}
	\draw (0,0) rectangle (8,.5);
	\filldraw[fill=gray] (0,0) rectangle (3,.5);
	
	\draw (0,0) rectangle (8,-.5);
	\filldraw[fill=gray] (0,0) rectangle (1,-.5); 
	\filldraw[fill=gray] (3,0) rectangle (5,-.5);
	
	\path (0,-1) -- (1,-1) node[midway,above]{$\Delta$};
	\path (1,-1) -- (3,-1) node[midway,above]{$d^\perp - \Delta$};
	\path (3,-1) -- (5,-1) node[midway,above]{$d^\perp - \Delta$};
	\path (5,-1) -- (8,-1) node[midway,above]{$n-2 d^\perp + \Delta$};
\end{tikzpicture}
\caption{Two codewords of weight $d^\perp$}
\label{fig:2-codewords}
\end{figure}
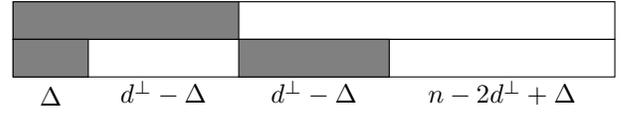

Each of the codewords covers exactly $d^\perp \binom{n-d^\perp}{i-1}$ $i$-sets. To calculate the total number of $i$-sets covered by these two codewords we need to subtract those $i$-sets that have been counted twice. They are of two kinds:

\begin{itemize}
\item Covered by the same pattern of size $i$ in $\cw h_1$ and $\cw h_2$. They have one position in the area of length $\Delta$ and all the other positions in the area of length $n-2 d^\perp + \Delta$. There are $\Delta \binom{n-2 d^\perp+\Delta}{i-1}$ such $i$-sets.

\item Covered by different patterns of size $i$ (at the same positions) in $\cw h_1$ and $\cw h_2$. They have one position in each of areas of length $d^\perp-\Delta$ and the remaining $i-2$ positions in the area of length $n-2 d^\perp + \Delta$. There are $(\Delta - d^\perp)^2 \binom{n-2 d^\perp+\Delta}{i-2}$ such $i$-sets.
\end{itemize}

Therefore these two codewords cover together the following amount of $i$-sets
\begin{multline*}
2 d^\perp \binom{n-d^\perp}{i-1} - \Delta \binom{n-2 d^\perp+\Delta}{i-1} \\
- (\Delta - d^\perp)^2 \binom{n-2 d^\perp+\Delta}{i-2} \; . 
\end{multline*}

This is at least $\mathfrak{M}(n, d^\perp, i)$.
We can now apply Theorem~\ref{thm:general-upper-bound} with $\tau=2$ and 
$u_i = \binom{n}{i} - \mathfrak{M}(n, d^\perp, i)$.
\end{IEEEproof}

It might be possible to further improve the bound in Corollary 2 by judiciously selecting 
three or more codewords in $\dual$, for example by taking three (or more) dual codewords of weight $d^\perp$. 
However, in that case it becomes more difficult to obtain good analytical estimates on $u_i$. Alternatively, 
it is also possible to choose some specific $\cw h_1, \cw h_2, \dotsc, \cw h_\tau$ and to 
compute all $u_i$ directly by computer. In that case, tighter bounds can be obtained. 
In the sequel, we refer to that method as a \emph{hybrid method}.


\section{Stopping Redundancy Hierarchy}

Consider a binary $[n, k, d]$ code $\code$. In Definition~\ref{def:stopping-redundancy} it is required that the stopping distance of the code defined by the parity-check matrix $H$ is $d$. However, a weaker requirement on the parity-check matrix of the code can be imposed. In this section, as it was suggested in~\cite{olgica2008permutation}, we require that the stopping distance of the code is at least $l$, for some $1 \le l \le d$. In that case, the number of rows in the parity-check matrix can be smaller than the stopping redundancy of the code.

\begin{define}[\hspace{-0.6ex} {\cite[Definition 2.4]{olgica2008permutation}}]
For $l \le d$, the $l$-th stopping redundancy of \code is the smallest nonnegative integer $\rho_l(\code)$ such that there exists a (possibly redundant) parity-check matrix $H$ of \code with $\rho_l(\code)$ rows and stopping distance at least $l$. The ordered set of integers
$\left(\rho_1(\code), \rho_2(\code), \dotsc, \rho_d(\code) \right)$
is called the \emph{stopping redundancy hierarchy} of \code. 
\end{define}

Note that the (conventional) stopping redundancy $\rho(\code)$ is equal to $\rho_d(\code)$. For codes with the minimum distance $d \ge 4$, neither two columns of the parity-check matrix are identical nor any of the columns equal to the all-zero vector. Therefore, $\rho_1(\code) = \rho_2(\code) = \rho_3(\code) = n-k$. Consequently, only $\rho_l(\code)$ for $l >3$ is of interest.

In \cite{olgica2008permutation}, the stopping redundancy hierarchy of binary linear codes is studied, and several upper bounds are obtained. 
In the sequel, we apply the ideas in previous section to the stopping redundancy hierarchy. 
We formulate a generalised version of Corollary~\ref{cor:2-rows}.

\begin{theorem}
\label{thm:hier-2-rows}
If $\dual$ contains at least two codewords of minimum weight $d^\perp$, then for $4 \le l \le d$,
\begin{equation*}
\rho_l(\code) \le 2 + \min_{t \ge r} \left\{ t + \kappa_t^{(1)}  \right\} + (r-l+1) \; . 
\end{equation*}
Moreover, if $(r-1)(l-1) \le 2^{l-1}$ then
\begin{equation*}
\rho_l(\code) \le 2 + \min_{t \ge r} \left\{ t + \kappa_t^{(2)}  \right\} \; , 
\end{equation*}
where
\begin{eqnarray*}
\kappa_t^{(i)} & = & \min \left\{k \in \Nat : Q_k(\lfloor \mathcal D_t^{(i)} \rfloor) = 0\right\} , \; i = 1, 2 \; , \\
Q_k(x) & = & P_k(P_{k-1}(\ldots P_1(x) \ldots)), \\
P_j(x) & = & \left\lfloor x \left( 1 - \frac{(l-1)2^{r-l+1}}{2^r-(2+t+j)} \right) \right\rfloor \; ,  \\
\mathcal D_t^{(1)} & = & \sum_{i=3}^{l-1} u_i \prod_{j=2}^{t+2}\left(1 - \frac{i \cdot 2^{r-i}}{2^r-j}\right) \; , \\ 
\mathcal D_t^{(2)} & = & \mathcal D_t^{(1)} +\frac{1}{2^{t-r}}\left(1+\frac{2/3}{2^{t-r+1}-1}\right) \; , \\
u_i &= & \binom{n}{i} - \mathfrak{M}(n, d^\perp, i) \; . 
\end{eqnarray*}
\end{theorem}
\begin{IEEEproof}
The case when $(r-1)(l-1) \le 2^{l-1}$ is analogous to the proof of Theorem~\ref{thm:general-upper-bound}, with the values of $\tau$ and $u_i$ as in Corollary~\ref{cor:2-rows}.

That proof, however, cannot be applied to the cases of small values of $l$ if the condition $(r-1)(l-1) \le 2^{l-1}$ does not hold. We note that this condition is required in the proof only to guarantee the uniform decrease of $\xi$ and $\eta$. Therefore, the argument for decrease of $\xi$ in the proof of Theorem~\ref{thm:general-upper-bound} can be applied as is. After that, we have to ensure that the constructed matrix is of the required rank $r$.

Note that since we have covered all the $i$-sets for $i=1,2,\dotsc, l-1$, the rank of the matrix is at least $l-1$. Hence, by adjoining at most $r-(l-1)$ rows, we finally obtain the required parity-check matrix.
\end{IEEEproof}

We note that tighter bounds on the stopping redundancy hierarchy could be obtained by using the hybrid method, discussed in the last paragraph of Section~\ref{sec:special-cases}.


\section{Numerical Experiments}

In this section, we compare the bounds on the stopping redundancy obtained in \cite{schwartz-vardy2006}, \cite{han2007improved}, \cite{han2008improved} with our results. We consider two codes: the extended $[24,12,8]$ binary Golay code and the extended $[48,24,12]$ binary Quadratic Residue (QR) code. Both of them are known to be self-dual (cf.\,\cite{houghten2003qr}).

The extended $[24,12,8]$ binary Golay code is arguably a remarkable binary block code. It is often used as a benchmark in studies of code structure and decoding algorithms. The code is self-dual, therefore $d^\perp = 8$. Moreover, it is known \cite[p.\,67]{macwilliams1977theory} that there are $759$ codewords of the minimum weight. The example of (conventional) parity-check matrix of the code is shown in Table \ref{tbl:Golay-matrix}, where the blank spaces denote zeroes.
In \cite{schwartz-vardy2006}, a greedy (lexicographic) computer search was used. It was found that the actual stopping redundancy of the extended $[24,12,8]$ binary Golay code is at most $34$.

\begin{table}
\caption{Parity-check matrix of the extended $[24,12,8]$ Golay code}
\label{tbl:Golay-matrix}
\[
	\left(
	\begin{smallmatrix}
	 1 & 1 & ~ & ~ & ~ & ~ & ~ & ~ & ~ & ~ & ~ & ~ & ~ & 1 & 1 & 0 & 1 & 1 & 1 & 0 & 0 & 0 & 1 & 0 \\
	 1 & ~ & 1 & ~ & ~ & ~ & ~ & ~ & ~ & ~ & ~ & ~ & ~ & 0 & 1 & 1 & 0 & 1 & 1 & 1 & 0 & 0 & 0 & 1 \\
	 1 & ~ & ~ & 1 & ~ & ~ & ~ & ~ & ~ & ~ & ~ & ~ & ~ & 1 & 0 & 1 & 1 & 0 & 1 & 1 & 1 & 0 & 0 & 0 \\
	 1 & ~ & ~ & ~ & 1 & ~ & ~ & ~ & ~ & ~ & ~ & ~ & ~ & 0 & 1 & 0 & 1 & 1 & 0 & 1 & 1 & 1 & 0 & 0 \\
	 1 & ~ & ~ & ~ & ~ & 1 & ~ & ~ & ~ & ~ & ~ & ~ & ~ & 0 & 0 & 1 & 0 & 1 & 1 & 0 & 1 & 1 & 1 & 0 \\
	 1 & ~ & ~ & ~ & ~ & ~ & 1 & ~ & ~ & ~ & ~ & ~ & ~ & 0 & 0 & 0 & 1 & 0 & 1 & 1 & 0 & 1 & 1 & 1 \\
	 1 & ~ & ~ & ~ & ~ & ~ & ~ & 1 & ~ & ~ & ~ & ~ & ~ & 1 & 0 & 0 & 0 & 1 & 0 & 1 & 1 & 0 & 1 & 1 \\
	 1 & ~ & ~ & ~ & ~ & ~ & ~ & ~ & 1 & ~ & ~ & ~ & ~ & 1 & 1 & 0 & 0 & 0 & 1 & 0 & 1 & 1 & 0 & 1 \\
	 1 & ~ & ~ & ~ & ~ & ~ & ~ & ~ & ~ & 1 & ~ & ~ & ~ & 1 & 1 & 1 & 0 & 0 & 0 & 1 & 0 & 1 & 1 & 0 \\
	 1 & ~ & ~ & ~ & ~ & ~ & ~ & ~ & ~ & ~ & 1 & ~ & ~ & 0 & 1 & 1 & 1 & 0 & 0 & 0 & 1 & 0 & 1 & 1 \\
	 1 & ~ & ~ & ~ & ~ & ~ & ~ & ~ & ~ & ~ & ~ & 1 & ~ & 1 & 0 & 1 & 1 & 1 & 0 & 0 & 0 & 1 & 0 & 1 \\
	 0 & ~ & ~ & ~ & ~ & ~ & ~ & ~ & ~ & ~ & ~ & ~ & 1 & 1 & 1 & 1 & 1 & 1 & 1 & 1 & 1 & 1 & 1 & 1 \\
	\end{smallmatrix}
	\right)
\]
\end{table}

It is known \cite[p.\,604]{macwilliams1977theory} that there are $17296$ codewords of the minimum weight in the extended $[48,24,12]$ binary Quadratic Residue (QR) code. The comparison of the upper bounds on the stopping redundancy is given in Table \ref{tbl:bounds}.

\begin{table}
\caption[Upper bounds]{Upper bounds on the stopping redundancy}
\label{tbl:bounds}
\centering
\begin{tabular}{lcc}
\hline\hline
 ~ & [24, 12, 8] Golay & [48, 24, 12] QR \\
\hline
 \cite[Thm 4]{schwartz-vardy2006} & {2509} & {4540385} \\
 \cite[Thm 1]{han2008improved} & 198 & {3655} \\
 \cite[Thm 3]{han2008improved} & 194 & {3655} \\
 \cite[Thm 4]{han2008improved} & 187 & {3577} \\
 \cite[Thm 7]{han2008improved} & 182 & {3564} \\
\hline
 Corollary \ref{cor:1-row} ($\tau=1$) & 180 & {3538} \\
 Corollary \ref{cor:2-rows} ($\tau=2$) & 177 & {3515} \\
\hline\hline
\end{tabular}
\end{table}

We also compare the bounds on stopping redundancy hierarchy in the previous chapter with the results for general codes, obtained in \cite{olgica2008permutation} (the bounds for cyclic codes therein are not applicable because neither of the codes is cyclic.) The numerical results are presented in Table \ref{tbl:hier-Golay-24} and Table \ref{tbl:hier-QR-48}.

\begin{table}
\caption{Bounds on the stopping redundancy hierarchy, $\rho_l$, for the extended $[24,12,8]$ Golay code}
\label{tbl:hier-Golay-24}
\centering
	\begin{tabular}{lcccc}
	\hline\hline
	$l$ & \cite[Thm 3.8]{olgica2008permutation} & \cite[Thm 3.11]{olgica2008permutation} & \cite[Thm 3.12]{olgica2008permutation} & Thm \ref{thm:hier-2-rows} \\
	\hline
	4 & 26  & 78   & --- & 25 \\ 
	5 & --- & 298  & --- & 36 \\ 
	6 & --- & 793  & 385 & 59 \\ 
	7 & --- & 1585 & --- & 103 \\ 
	8 & --- & 2509 & --- & 177 \\ 
	\hline\hline
	\end{tabular}
\end{table}

\begin{table}
\caption{Bounds on the stopping redundancy hierarchy, $\rho_l$, for the extended $[48,24,12]$ QR code}
\label{tbl:hier-QR-48}
\centering
	\begin{tabular}{lccc}
	\hline\hline
	$l$ & \cite[Thm 3.8]{olgica2008permutation} & \cite[Thm 3.11]{olgica2008permutation} & Thm \ref{thm:hier-2-rows} \\
	\hline
	4  & 42  & 300           & 47 \\ 
	5  & 62  & 2 324          & 58 \\ 
	6  & 105 & 12 950         & 92 \\ 
	7  & --- & 55 454         & 158 \\ 
	8  & --- & 190 050        & 287 \\ 
	9  & --- & 536 154        & 514 \\ 
	10 & --- & 1 271 625 & 978 \\ 
	11 & --- & 2 579 129 & 1856 \\ 
	12 & --- & 4 540 385 & 3515 \\ 
	\hline\hline
	\end{tabular}
\end{table}

\medskip
Next, we use the hybrid method, mentioned in the last paragraph of Section~\ref{sec:special-cases}. We take $\tau$ first rows of conventional parity-check matrix of the extended $[24,12,8]$ Golay code (Table~\ref{tbl:Golay-matrix}), for $1 \le \tau \le 12$, compute all $u_i$, and apply techniques similar to Theorem~\ref{thm:general-upper-bound} and Theorem~\ref{thm:hier-2-rows}. Numerical results are presented in Table~\ref{tbl:hier-hybrid-Golay-24}.

\begin{table}
\caption{{Bounds on the stopping redundancy hierarchy, $\rho_l$, derived by the hybrid method for the extended $[24,12,8]$ Golay code}}
\label{tbl:hier-hybrid-Golay-24}
\centering
    \begin{tabular}{c|ccccc}
    \hline\hline
    $\rho_l$ & $l = 4$ & $l = 5$ & $l = 6$ & $l = 7$ & $l = 8$ \\
    \hline
     $\tau=1$ & 24 & 36 & 61 & 105 & 180 \\
     $\tau=2$ & 24 & 36 & 59 & 103 & 177 \\
     $\tau=3$ & 25 & 35 & 58 & 102 & 175 \\
     $\tau=4$ & 25 & 34 & 57 & 100 & 174 \\
     $\tau=5$ & 26 & 33 & 56 & 99 & 172 \\
     $\tau=6$ & 27 & 33 & 56 & 98 & 171 \\
     $\tau=7$ & 28 & 33 & 55 & 98 & 170 \\
     $\tau=8$ & 29 & 33 & 55 & 97 & 169 \\
     $\tau=9$ & 30 & 33 & 55 & 96 & 168 \\
     $\tau=10$ & 31 & 33 & 55 & 96 & 167 \\
     $\tau=11$ & 32 & 34 & 55 & 96 & 167 \\
     $\tau=12$ & 33 & 35 & 56 & 97 & 168 \\
     \hline\hline
    \end{tabular}
\end{table}


\section{Acknowledgment}

The authors wish to thank  {\O{}}yvind Ytrehus for helpful discussions. 


\bibliographystyle{IEEEtran}
\bibliography{IEEEabrv,current-notes}

\end{document}